# Pressure-induced Structure Change and Anomalies in Thermodynamic Quantities and Transport Properties in Liquid Lithium Hydride

*Xiao Z. Yan* [1,2,4], *Yang M. Chen*[2], *Hua Y. Geng**[1,3], *Yi F. Wang*[1], *Yi Sun*[1], *Lei L. Zhang*[1], *Hao Wang*[1] *and Yin L. Xu*[1]

1. National Key Laboratory of Shock Wave and Detonation Physics, Institute of Fluid Physics, CAEP, P.O. Box 919-102, Mianyang 621900, Sichuan, People's Republic of China
2. School of Science, Jiangxi University of Science and Technology, Ganzhou 341000, Jiangxi, People's Republic of China
3. HEDPS, Center for Applied Physics and Technology, and College of Engineering, Peking University, Beijing 100871, People's Republic of China
4. Jiangxi Provincial Key Laboratory of Particle Technology, Ganzhou 341000, Jiangxi, People's Republic of China

Abstract

Understand the nature of liquid structure and its evolution under different conditions is a major challenge in condensed physics and materials science. Here, we report a pressure-induced structure change spanning a wide pressure range in liquid-state lithium hydride (LiH) by first-principles molecular dynamic simulations. This behavior can be described as a continuous crossover from low pressure liquid with $Li^+$-$H^-$ duality symmetry to high pressure one with broken of duality symmetry. The thermodynamic quantities such as heat capacity and ionic transport properties such as diffusivity are also saliently impacted. It is important to stress that such behavior is firstly predicted for this category of materials, which is ubiquitous in universe as well as in industry applications. Lastly, a comprehensive high-pressure high-temperature phase diagram of LiH is constructed, which embodies rich physics in this previously-thought-simple ionic compound.

---

* Author to whom correspondence should be addressed: s102genghy@caep.cn.

Due to the lack of symmetry and periodicity of atomic arrangement, there is no ideal model to describe the structure of liquid. Understanding the liquid structure and property has been one of the most challenging problems in condensed matter physics and materials science [1-3]. Similar to the crystalline one, the atomic structure and properties of liquid may experience drastic change under certain conditions, leading to the occurrence of liquid-liquid phase transitions (LLPT).

In line with intuitional comprehension, LLPT has mostly been (but not always) suggested in molecular or atomic liquids that with some covalent characteristics. Typical examples include the liquid carbon [4], wherein a dominantly $sp^2$-coordinated liquid changes to a denser, mainly $sp^3$-coordinated one; and the liquid phosphorus [2, 5], which transforms from a molecular liquid comprising tetrahedral $P_4$ molecules to a polymeric form. The transition in liquid silicon [6, 7] and germanium [8] is similar to that of carbon and phosphorus. Other examples include the transition and dissociation in hydrogen from molecular $H_2$ to atomic H [9-12], and the transition in water from a low-density form with an open, hydrogen-bonded tetrahedral structure, to a high-density form with a collapsed second coordination shell [13-18]. Besides, LLPT has also been reported to occur in melts of pure metal Ce [19], Ga [20], Li [21-23] and K [24], and alloys of $Al_2O_3$–$Y_2O_3$ [25] and $La_{50}Al_{35}Ni_{15}$ [26]. The structure changes in these later systems primarily originate from either strong-correlations of $f$ electrons or residual covalent bonding arisen from core-core electron interactions and valence electron localization or delocalization that leading to hybridizations.

For ionic melt, due to the dominance of electronic Coulomb interactions, the structure is simple, and the influence of pressure or temperature on atomic bonding (as well as the electronic structure) is less obvious than that in other systems. By far, there is no LLPT has been reported in this category of system to the best of our knowledge.

LiH is the simplest ionic compound, even though its elemental constituents show complex behavior at high pressures [27-29]. Under ambient conditions, LiH crystallizes in the six-fold coordinated NaCl (B1) structure [30]. Similar to the alkali halides (and other simple salts with B1 structure), a phase transition from B1 to another high-symmetric eight-fold coordinated CsCl (B2) structure has been predicted theoretically at about 300 GPa [31-34], and the latter is stable beyond 1000 GPa. Experimental signatures also suggested that the B1-B2 structure transition accompanied with metallization might be achieved when beyond 252 GPa [35].

The behavior of liquid LiH structure is also expected to be akin to that of alkali halides, which have a duality symmetry between the cations and anions. Ogitsu *et al.* have carried out first-principles molecular dynamic (MD) simulations using density functional theory (DFT) and Car-Parrinello (CP) method [36] to study the melting of LiH under pressure [37]. They found that the variation of melting curve is similar to that of alkali halides. And the compressed liquid close to the melting line still retains the ionic character both at low and high pressures. However, when comparing with experimental data [38, 39], it is found that their data underestimate the melting temperature by 18% and 20% at 0 GPa and 4 GPa, respectively. This big discrepancy casts doubt on the applicability of DFT in LiH, which however should be appropriate if considering from the perspective of physics.

In this letter, we revisit this compelling problem by performing much more careful and deliberate first-principles MD simulations using Born-Oppenheimer (BO) method [40] (rather than CP dynamics) based on DFT to study the melting, the liquid structure, and the properties of LiH. Our results not only reconcile the aforementioned huge discrepancy, but also reveal exotic behaviors of structure, thermodynamic quantities and ionic transport properties in this simple ionic compound.

The BO-MD simulations are performed within the Vienna ab initio simulation package (VASP) [41]. The electron-ion interaction is described by the projector augmented-wave (PAW) pseudopotential [42], and the Perdew-Burke-Ernzerhof (PBE) [43] parametrizations for the electron exchange-correlation energy functional are employed. The hard version of the PAW pseudopotentials provided in VASP code are used. The energy cutoff for the plane-wave basis set is 700 eV. The convergence of the k-point sampling meshes is carefully checked, which gives an error in pressure and energy better than 0.5 GPa and 5 meV/atom, respectively. Other computational details are given in the Supporting Information [44].

In order to determine the melting curve of LiH, we simulate the melting of both B1 and B2 phases firstly by using the "Z-curve" method [45] with the microcanonical (NVE) ensemble. The obtained melting temperatures as a function of pressure in comparison with the experimental data and Ogitsu's two-phase method calculation are shown in Fig. 1. In contrast to the big underestimation in the results of Ogitsu *et al.*, the Z-method slightly overestimates the melting temperatures. For example, at 0 GPa and 4 GPa, the melting temperatures are 1040 K and 1430 K, respectively, which overestimate by about 10% against the experimental data [38, 39].

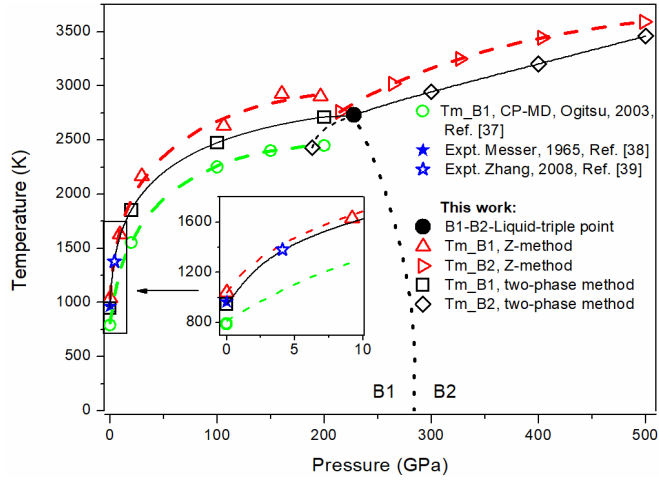

Fig. 1. Melting curves of LiH in B1 and B2 phases. The solid and dashed lines are obtained by using a Kechin fit of the respective simulated melting temperatures ($T_m$). The fitting parameters are shown in Table S1 in Supporting Information [44]. Inset: An enlarged view of the region of 0-10 GPa. The dotted line indicates the B1-B2 phase boundary predicted by Ref. [34].

It should be noted that though the Z-method works well in predicting the melting of $H_2$ and metallic H [29, 46], it was also reported that sometimes it could overestimate the melting temperatures [47, 48] when the solid is harmonic. We checked this in LiH and found that the anharmonicity is small in both B1 and B2 phase. In order to obtain a more reliable melting curve, we then simulate the melting using two-phase method with the isothermal-isobaric (NPT) ensemble. Our two-phase simulations give a melting temperature of 955 K at 0 GPa, which is in a perfect agreement with experimental data (948 K ~ 973 K) [38, 39]. Although we did not simulate the melting at 4 GPa directly, a Kechin [49] fit of our two-phase melting data predicts the melting temperature to be 1377 K, which also agrees well with the experiment (1380 K) [39]. These new theoretical data reconcile the DFT-experiment discrepancy. The deviation between our results and Ogitsu's may be due to that the electrons in their CP-MD are close to, but not exactly on the Born-Oppenheimer surface, which would lead to an underestimation of the energy and pressure. Furthermore, all the predicted melting curves behave similarly to alkali halides [50, 51], *i.e.*, it is steep at 0-100 GPa and becomes flat when above this pressure.

Fig. 2 displays the pair correlation functions (PCFs) $g_{i\text{-}j}(r)$ of compressed liquid LiH close to and above the melting curve, the evolution of which provides overall information on the structure change. At low pressures, *e.g.*, at 7 GPa as plotted in the bottom of Fig. 2(a), the PCFs reveal a structural feature analogous to typical molten alkali halides [50-52]. The $g_{\text{Li-H}}(r)$ exhibits a characteristic ionic peak, and is antiphase to $g_{\text{H-H}}(r)$ and $g_{\text{Li-Li}}(r)$. Namely, the maxima of $g_{\text{Li-H}}(r)$ are located at the minima of the latter two which nearly overlap with each other. This feature implies an alternating but identical distribution of cation ($Li^+$) and anion ($H^-$) shells in the low-pressure liquid LiH, which possessing a dual symmetry between the cation and anion sub systems. This similarity in structure of LiH to the molten alkali halides is expectable, since the nature of hydrogen at low pressure is similar to halogens with a high electronegativity.

On increasing the temperature further, $g_{\text{H-H}}(r)$ and $g_{\text{Li-Li}}(r)$ become more distinguishable, and the $Li^+$ and $H^-$ dual symmetry starts to break. At 2586 K and 11 GPa (the bottom of Fig. 2(b)), the first peak of $g_{\text{H-H}}(r)$ reduces more when compared to that of $g_{\text{Li-Li}}(r)$. Furthermore, the $g_{\text{H-H}}(r)$ is characterized by the appearance of a prepeak located at 0.8 Å, which suggests possible instant $H_2$ units (it should be interpreted as random encounters of H anion rather than a chemical species due to the very short lifetime of less than 8 fs). It is thus conceivable that thermal excitation drives some of $H^-$ anions to break through the repulsion shell of $Li^+$ cations.

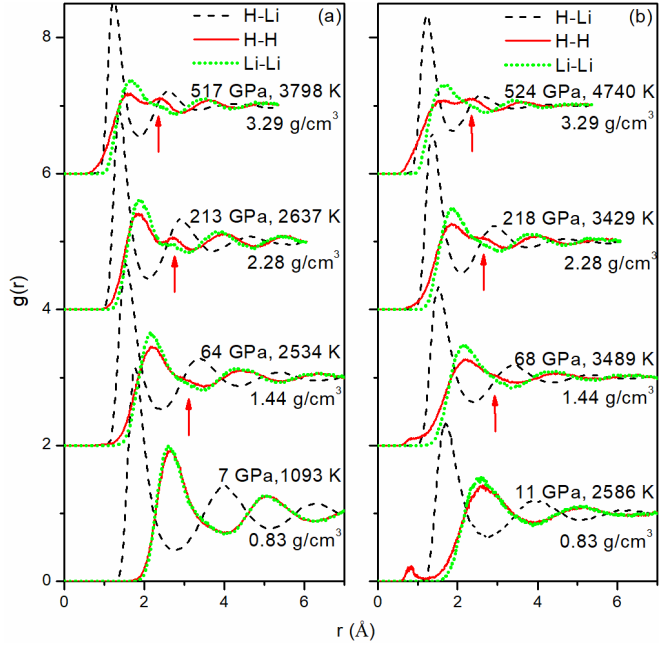

Fig. 2. Calculated pair correlation functions (PCF) of liquid LiH for (a) close to and (b) far above the melting curve. The gradually developing of a new main peak is assigned by arrows. PCFs for different thermodynamic conditions are vertically shifted for a succinct display purpose.

As pressure increases, the position of the first peak of $g_{Li-H}(r)$ shows a weak pressure dependence, whereas for $g_{H-H}(r)$ and $g_{Li-Li}(r)$, the pressure dependences are much more prominent, and the breaking of $Li^+$-$H^-$ duality symmetry becomes evident. We also notice that as pressure increases from 7 GPa to 213 GPa along the isotherm at 2600 K, the averaged nearest neighboring (NN) H-Li distance decreases from 1.75 Å to 1.38 Å (of 0.37 Å, or 21% shrinkage), whereas the averaged NN H-H and Li-Li distance both decrease from 2.62 Å to 1.85 Å (of 0.77 Å, or 29% shrinkage). This result indicates that the compression in LiH liquid is more endured by the space between the H-Li ion pairs, rather than the intra-distance between them, which bears a resemblance to a molecular liquid.

More importantly, a remarkable characteristic of the compressed liquid is the gradually developing of a new main peak locating at the high distance side of the first main peak of $g_{H-H}(r)$ when P > 50 GPa. As shown in Fig. 2, this peak has a magnitude comparable to the first main peak at ~500 GPa, and overtakes the latter at higher pressures. After the emergence of this new peak, the maxima of $g_{H-H}(r)$ gradually dephases from that of $g_{Li-Li}(r)$, and approaching an in-phase variation with $g_{Li-H}(r)$. This is in stark contrast to the low-pressure liquid, implying that H atoms already penetrate into the first H-Li coordination shell in the high-pressure liquid phase, and the $Li^+$-$H^-$ duality symmetry is completely broken. By comparing the $g_{H-H}(r)$ of solid B1 and B2 with that of liquid, we find that the H-H local structure of high-pressure liquid is closely related to the B2 phase (see Fig. S6 in the Supporting Information [44]).

The coordination number (CN) estimated by integrating $g_{i-j}(r)$ over the first peak also supports this argument. It is known that the CN of B1 and B2 structure is 6 and 8, respectively. Our results reveal that the CNs of H-Li (H atoms around each Li) and H-H in liquid LiH increase from 4.5 to 7 and from 13 to 22, respectively, when the pressure increases from 0 to 200 GPa (Fig. S7). These provide evidence for local structure change in compressed liquid LiH.

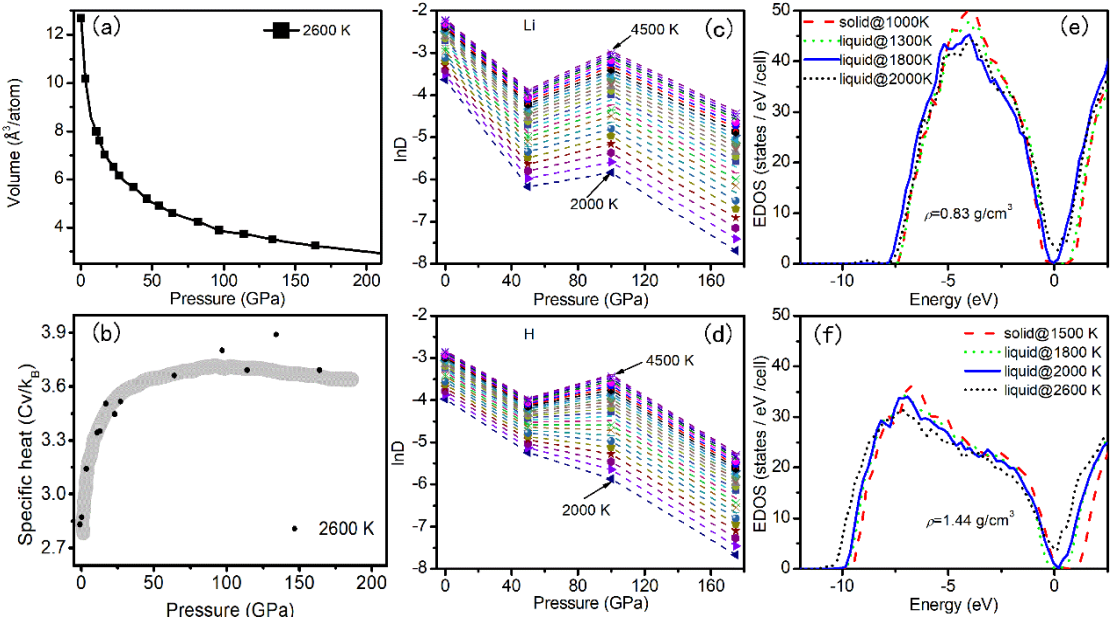

Fig. 3. (a) Isothermal pressure-volume P(V) curves of liquid LiH. (b) Specific heat extracted from thermal fluctuations in BO-MD simulations at the given temperatures by $k_BT^2C_v = \langle(E-\langle E \rangle)^2\rangle$, where $\langle \cdots \rangle$ denotes ensemble average. The bold line is for the eye-guide only. (c and d) The natural logarithm of self-diffusion coefficients at different temperatures as a function of pressure. (e and f) The calculated electronic density of states (by averaging twenty snapshots of the MD trajectories) at 0.83 g/cm$^3$ (corresponding to the density under ambient condition) and 1.44 g/cm$^3$ (corresponding to a pressure of 50 GPa at 0 K), respectively.

The equation of state P(V) at 2600 K is plotted in Fig. 3(a), where no perceptible volume discontinuity is observed, suggesting the structure change in liquid LiH is not a first-order phase transition. To verify whether it is a higher order transition or not, we then calculate the higher derivatives of free energy, *i.e.,* the heat capacity. According to the fluctuation-dissipation theorem, fluctuations in energy give the specific heat. As presented in Fig. 3(b), our results show that the pressure-dependent specific heat displays a continuous increase across the structure transition region. Therefore, this behavior can be described as a pressure-induce crossover from low-pressure duality symmetry liquid to high- pressure broken duality symmetry liquid.

The local structure change in liquid could modify its transport properties. We compute the self-diffusion coefficients of liquid LiH from the long-time limit of the slope of the mean square displacements (MSD). The temperature- and pressure-dependent natural logarithm of self-diffusion coefficients of H and Li atoms are shown in Fig. 3(c-d). It is shown that for a given pressure, the self-diffusion coefficients increase with temperature. This is common in liquid since the atoms become more mobile under higher temperature. However, the pressure dependent self-diffusion coefficients for Li at all temperatures considered here and for H at higher temperatures exhibit abnormal increase in 40-100 GPa pressure range. This abnormal increase of diffusivity can be explained by the broken of dual symmetry of $Li^+$-$H^-$ which weakened the bonding of $Li^+$ and $H^-$.

Previous studies suggested that the solid B1 phase remain insulating until it transforms to B2 phase under high pressure [34, 35]. In Fig. 3(e) and (f), we plot the calculated electronic density of states for LiH under high temperatures. It is found that both the solid (B1) and liquid phases near the melting curve have a finite energy gap when P < 50 GPa. In other words, when solid (B1) LiH melts at low pressure region, it firstly enters an insulating state. This insulator liquid then will metalize at a higher temperature. It should be noted that our results are based on the calculations using PBE exchange-correlation functional, which always underestimates the band gap of a material. It is known that the HSE functional [53] can improve the gap problem greatly. Due to the high computational demanding HSE, we employed a relatively smaller supercell containing 64 atoms to estimate the gap underestimation by PBE against HSE. The results (Fig. S11) show that PBE underestimates the gap by 0.8 eV compared to HSE06 functional [53]. Therefore, the actual temperature-induced insulator-metal transition line should correspondingly be shifted up to a higher temperature compared to that of the PBE result as shown here (Fig. 4).

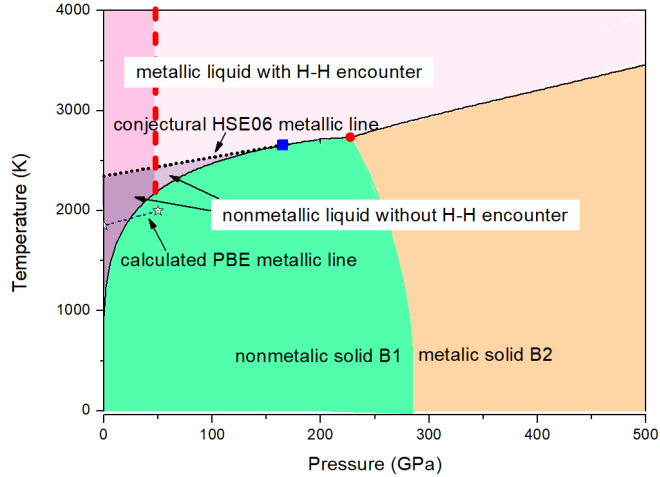

Fig. 4. Phase diagram of LiH. The solid lines are melting curves of B1 and B2 phases obtained by the two-phase simulations. The upright bold dashed line sketches the crossover from low-pressure liquid with $Li^+$-$H^-$ duality symmetry to high-pressure liquid with broken of duality symmetry estimated by tracing the structure changes and specific heat jump. The dotted lines represent the estimated metallization lines at PBE and HSE06 level, respectively. The liquid with instant H-H encounter and that without H-H encounter are determined by mapping over the characteristic $H_2$ prepeak in the PCFs (Fig. S8). The solid B1 and B2 phase boundary is obtained by quasi-harmonic approximation in our previous work [Ref. 34]. The filled circle represents the B1-B2-liquid triple point at about 228 GPa and 2730 K. The filled square indicates the HSE06 estimated B1-nonmetal liquid-metal liquid triple point at about 166 GPa and 2655 K.

All of these discoveries sum up to a completely new and comprehensive phase diagram of LiH, as depicted in Fig. 4. There are two triple points being identified, two kinds of metallization transition are reported (one is in solid B1-B2 transition, the other is in nonmetallic-metallic liquid transition), and a pressure-induced liquid crossover is documented. Unambiguous numerical evidences for liquid structure transition are provided by state-of-the-art first-principles MD simulations. We don't rule out the existence of LLPT with the critical point locating deeply within the B1 solid phase. We tried to explore this critical point by using BO-MD. But it failed for the difficulty to get a supercooled liquid LiH, possibly because our employed supercell is not large enough for this purpose.

The new phase diagram of LiH is much more complex than previously thought, and contains rich physics for the liquid of this simple salt. Since LiH shares common electronic and atomic structures and features with many other simple ionic compounds such as NaCl, LiF and MgO etc. at ambient conditions, the discovered pressure-driven structure change and associated anomalies in thermodynamic quantities and transport properties might be general for this category of important materials that have cation-anion duality symmetry at ambient pressure.

## ACKNOWLEDGEMENTS

This work is supported by the National Natural Science Foundation of China under Grant Nos. 12364003, 11672274, 11704163 and 11804131, the NSAF under Grant No. U1730248, the Science challenge Project under Grant No. TZ2016001, and the CAEP Research Project CX2019002, the Natural Science Foundation of Jiangxi Province under Grant Nos. 20232BAB211022 and 20181BAB211007, the Natural Science Foundation of Henan Province under Grant No. 242300421689, the Educational Commission of Jiangxi Province of China under Grant No. GJJ200862.

# Supplementary Information for "Pressure-induced Structure Change and Anomalies in Thermodynamic Quantities and Transport Properties in Liquid Lithium Hydride"

*Xiao Z. Yan* [1, 2, 4], *Yang M. Chen*[2], *Hua Y. Geng**[1, 3], *Yi F. Wang*[1], *Yi Sun*[1], *Lei L. Zhang*[1], *Hao Wang*[1] *and Yin L. Xu*[1]

1. National Key Laboratory of Shock Wave and Detonation Physics, Institute of Fluid Physics, CAEP, P.O. Box 919-102, Mianyang 621900, Sichuan, People's Republic of China
2. School of Science, Jiangxi University of Science and Technology, Ganzhou 341000, Jiangxi, People's Republic of China
3. HEDPS, Center for Applied Physics and Technology, and College of Engineering, Peking University, Beijing 100871, People's Republic of China
4. Jiangxi Provincial Key Laboratory of Particle Technology, Ganzhou 341000, Jiangxi, People's Republic of China

**Supplementary Information**

**A. Melting:**

In order to get a reliable melting temperature of LiH under compression, we simulate the melting of solid B1 and B2 phases by employing both the Z-method and two-phase method. The Z-method simulations are performed using the microcanonical (NVE) ensemble, in which the particle number N, internal energy E, and cell volume V are conserved quantities. The Baldereschi mean value point [1] is used to sample the first Brillouin zone, the convergence of which is tested by using a 4 × 4 × 4 Monkhorst-Pack grid. The size of the simulation supercell contains 216 and 250 atoms for B1 and B2 phase, respectively, with periodic boundary conditions being imposed. The size of the time-step for integration of the classical motion equations is 0.5 fs. A typical MD simulation runs 5000 time steps, corresponding to 2.5 ps. The structure equilibrating and melting usually take place within less than 1 ps in the MD simulations. The equilibrium pressure and temperature are obtained by statistical averaging over the last 2000 time steps. For a series of NVE simulations, we gradually increase the energy by adjusting the initializing temperature for every fixed V. Along this isochoric line (see Fig. S1), the solid phase first evolves into a superheated state, and then abruptly collapses to a liquid state after reaching a critical point. The corresponding thermodynamic condition after the structural collapse gives exactly the melting pressure and temperature.

* Author to whom correspondence should be addressed: s102genghy@caep.cn.

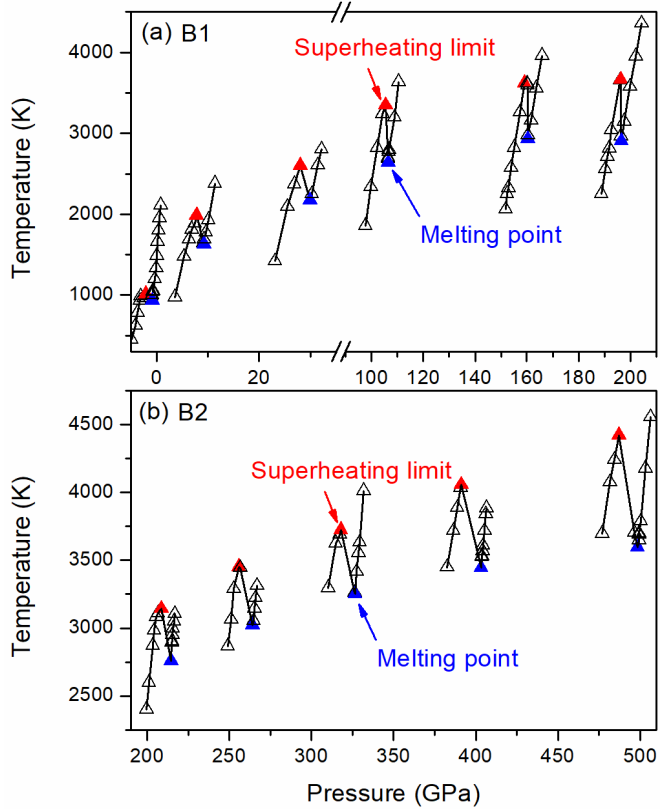

**Fig S1.** Typical Z-curves in the Z-method calculation of the melting of (a) B1 and (b) B2 phase of LiH.

In the simulations of two-phase method, the isothermal-isobaric ensemble (NPT, N is the particle number, P is the pressure and T is the temperature) is employed. The simulation cell contains 432 and 500 atoms for B1 and B2 phase, respectively, with periodic boundary conditions being imposed. The integration over the first Brillouin zone is carried out using a $2 \times 2 \times 1$ k-point grid. Denser grid of $3 \times 3 \times 2$ is also used to check the convergence, wherein the variation in pressure and energy are less than 0.5 GPa and 2 meV/H, respectively (see Fig. S2). To achieve an NPT ensemble, the Parrinello-Rahman dynamics with Langevin thermostat is used. The fictitious mass and friction coefficient for lattice degrees of freedom are set to 40 atomic mass units and 50 ps$^{-1}$, respectively; and the friction coefficient for atomic degrees of freedom used in Langevin dynamics is set as 50 ps$^{-1}$. This setting leads to an efficient coupling with the thermostat and barostat, and equilibrates the system within less than 1 ps, as indicated in Fig. S3. This good quality of MD simulations ensures a reliable two-phase modeling of the melting with NPT ensemble. Typical configurations of two-phase equilibration are illustrated by the snapshots shown in Fig. S4. It is necessary to point out that for the cell size we employed here, it is very difficult to maintain the two-phase equilibrating coexistence interface for a long time. Usually, the system will evolve towards the solid or liquid phase quickly. Nonetheless, the uncertainty in the melting temperature introduced by this feature is less than 50 K, which is good enough for our purpose.

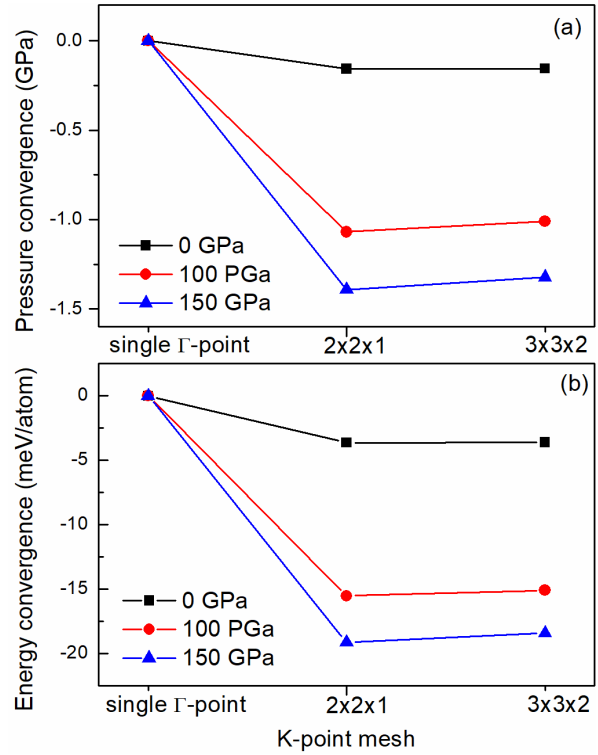

**Fig. S2.** Convergence test of k-point mesh in (a) pressure and (b) energy, respectively. The data of different pressure are shifted to align at the single Γ-point.

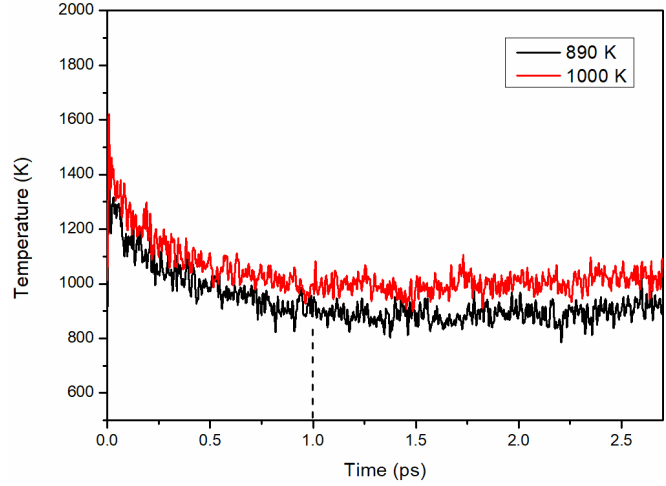

**Fig. S3.** Typical relaxation of the temperature in MD simulations for modelling the structure equilibrating in the solid B1 phase at about 0 GPa with an NPT ensemble at 890 K and 1000 K, respectively. The system achieves an equilibrium state within 1 ps.

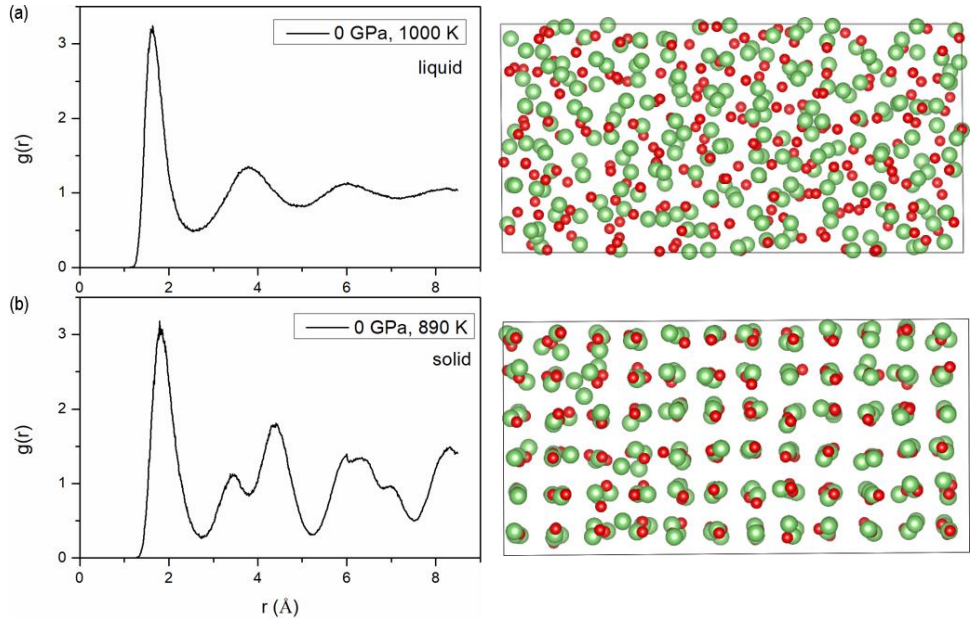

**Fig. S4.** (Left panels) The pair correlation functions *g(r)* and (Right panels) the corresponding MD snapshots of equilibrating LiH at a pressure of 0 GPa, and a temperature of 1000 K (liquid) and 890 K (solid) with two-phase method, respectively.

**Table S1.** Fitting of the melting temperatures to Kechin equation of $T_m = T_0 (1 + \Delta P/a)^b \exp(-c\Delta P)$. The negligible parameter c indicates that there is no anomaly in the melting curve at higher pressures, and the Kechin equation reduces back to Simon

equation of $T_m = T_0 (1 + \Delta P/a)^b$.

|  | phase | a (GPa) | b | c (GPa$^{-1}$) |
|---|---|---|---|---|
| Two-phase method | B1 | 0.80208 | 0.20967 | 0.0005 |
|  | B2 | 1.43362 | 0.02713 | -0.0007 |
| Z-method | B1 | 1.02927 | 0.24616 | 0.0008 |
|  | B2 | 136.6992 | 0.38188 | 0.0006 |

**B. Liquid structure and properties:**

In order to explore the structure and property of LiH across a wide range of pressure and temperature, we further simulate the liquid behavior with the canonical (NVT) ensemble, where the particle number N, cell volume V and temperature T are conserved quantities. The Baldereschi mean value point is used to sample the first Brillouin zone, which gives a good description of the total energy and pressure, with an accuracy comparable to a regular 4 × 4 × 4 meshes. The size of the simulation cell contains 216 and 250 atoms for B1 and B2 as the starting configuration, respectively, with periodic boundary conditions being imposed. The time step for integration of the classical motion equations is 0.5 fs. A typical MD simulation runs at least 10000 time steps, corresponding to 5 ps. The structure and properties are obtained by statistical average over the last 5000 time steps. Long-time MD simulations with more than 25 ps have also been performed to examine the convergence of pair correlation functions and specific heat (see Fig. S5).

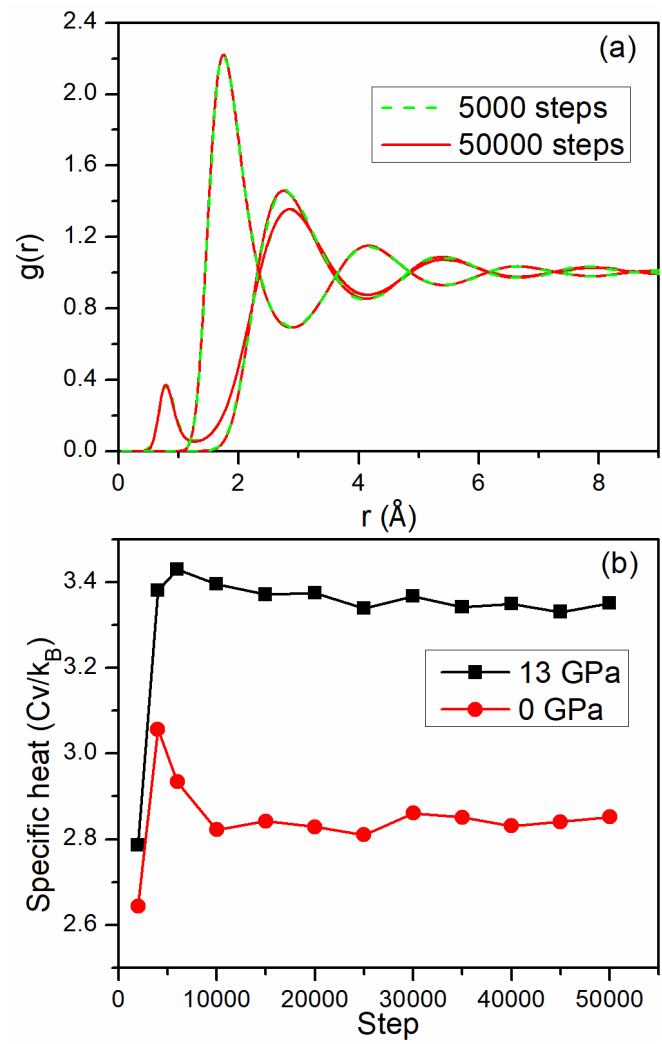

**Fig. S5.** Convergence test of (a) pair correlation functions and (b) specific heat.

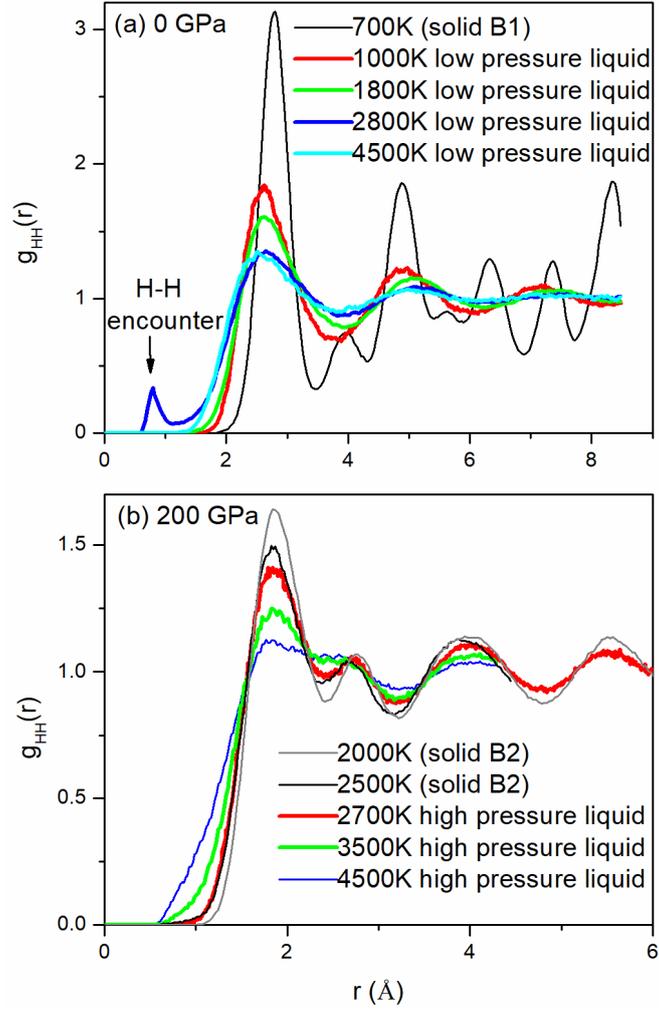

**Fig. S6.** Comparison of H-H pair correlation functions of solid B1 and B2 with liquid at (a) 0 GPa and (b) 200 GPa. The results indicate that the structure of high-pressure liquid phase is closely related to the solid B2 phase, which might be the driven force to break the cation-anion dual symmetry in liquid LiH. Note: there is no superionic state observed in both B1 and B2 phase.

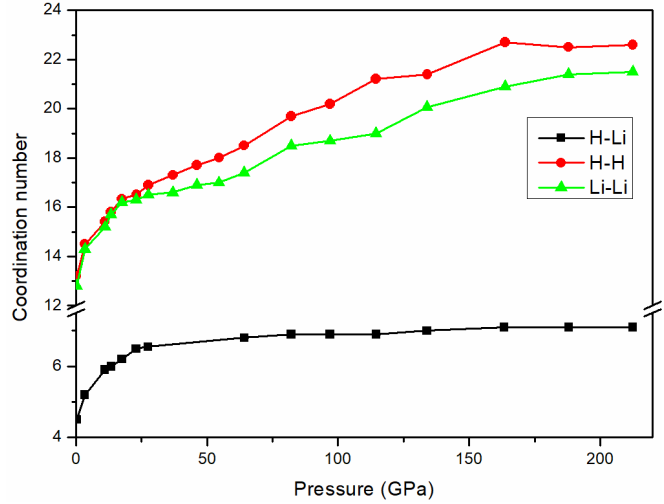

**Fig. S7.** Variation of the coordination number (CN) with pressure in liquid LiH at 2600 K. Note that the H-H and Li-Li CNs diverge with each other at about 25 GPa and saturate after 150 GPa, whereas H-Li CN saturates at about 25 GPa.

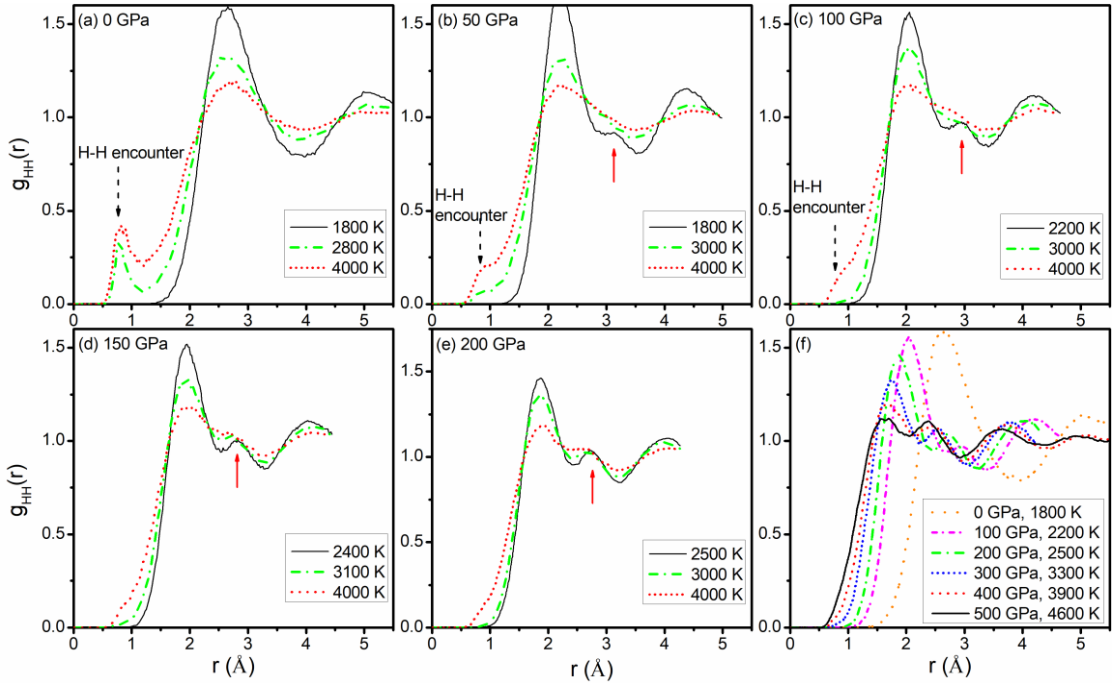

**Fig. S8.** (a-e) Evolution of H-H pair correlation function with temperature at a given pressure; (f) Variation of H-H pair correlation function with both temperature and pressure simultaneously. The solid arrows indicate the distinct features for the liquid-liquid transformation.

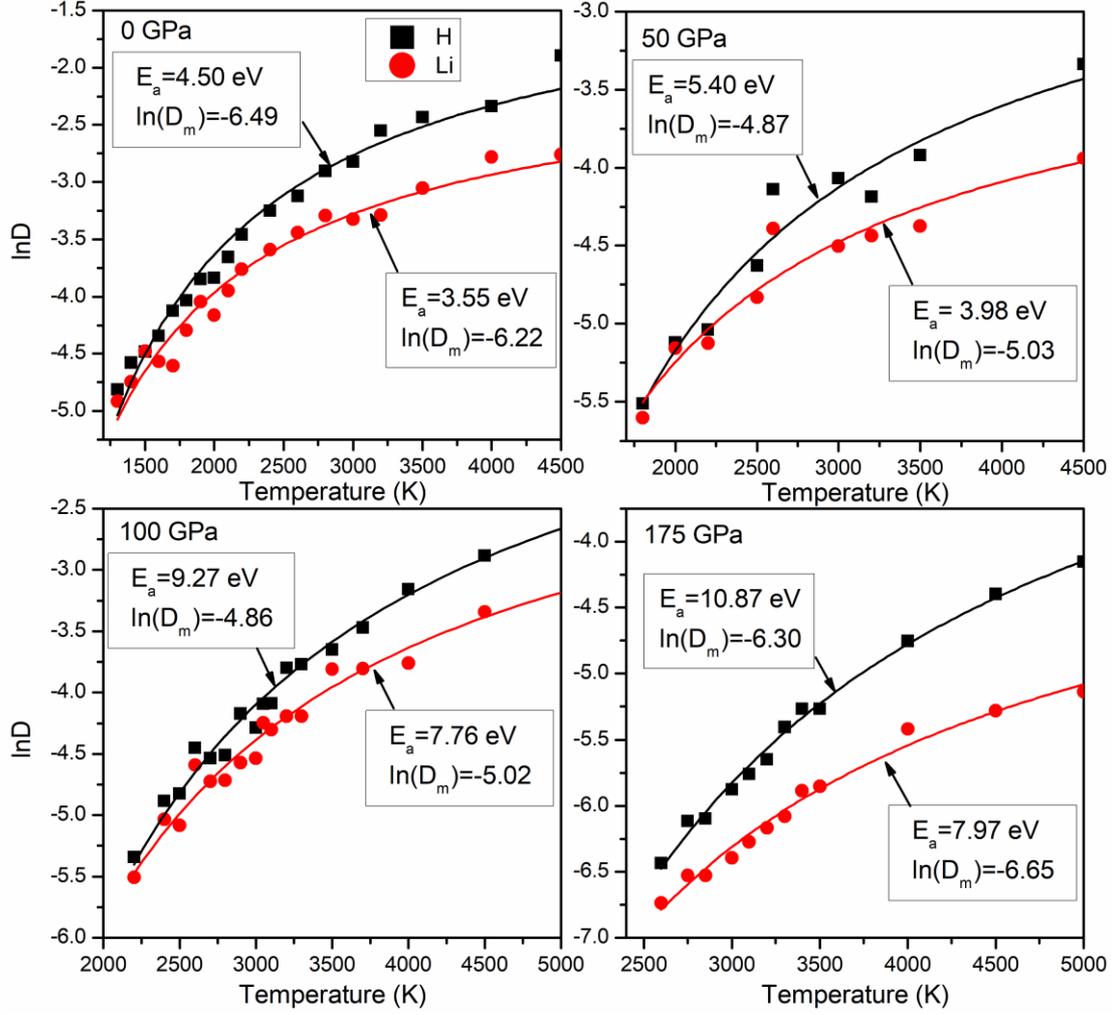

**Fig. S9.** The temperature dependent natural logarithm of self-diffusion coefficients of H (squares) and Li (cycles) at different pressures. The calculated data are fitted to a formula of $\ln\frac{D}{D_m} = -\frac{E_a}{k_B T} + \frac{E_a}{k_B T_m}$, in which $E_a$ is the activation energy, $T_m$ is the melting temperature, $D_m$ is the self-diffusion coefficient at the melting temperature and $k_B$ is the Boltzmann constant.

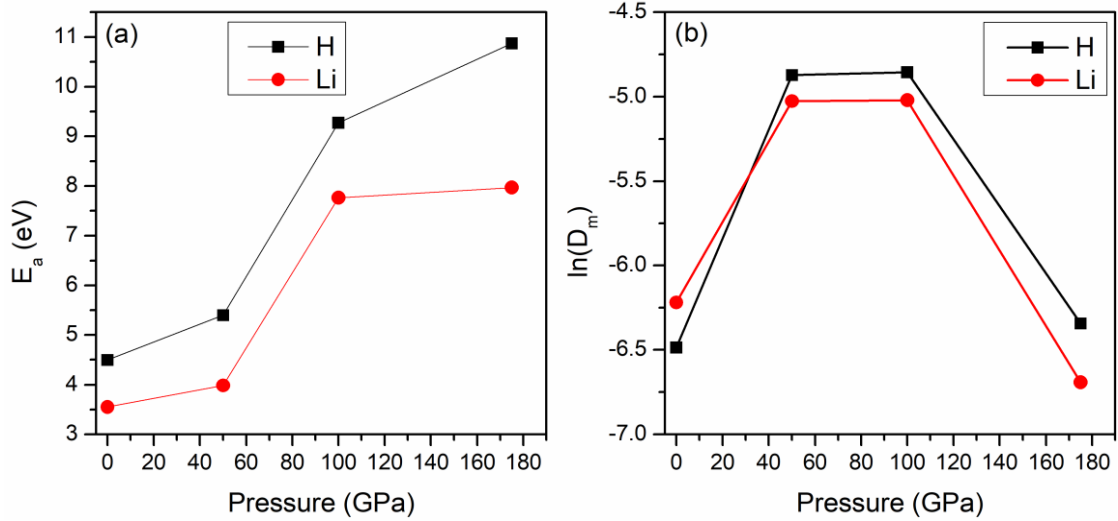

**Fig. S10.** (a) Variation of the estimated diffusion activation energy with pressure, and (b) variation of the estimated self-diffusion coefficients of H and Li at the corresponding melting temperature with pressure. The lift of the activation energy with pressure reflects the greater confinement of the ions in the compressed liquid LiH. The anomalous change in $E_a$ and $D_m$ is in accordance with the pressure-induced liquid crossover.

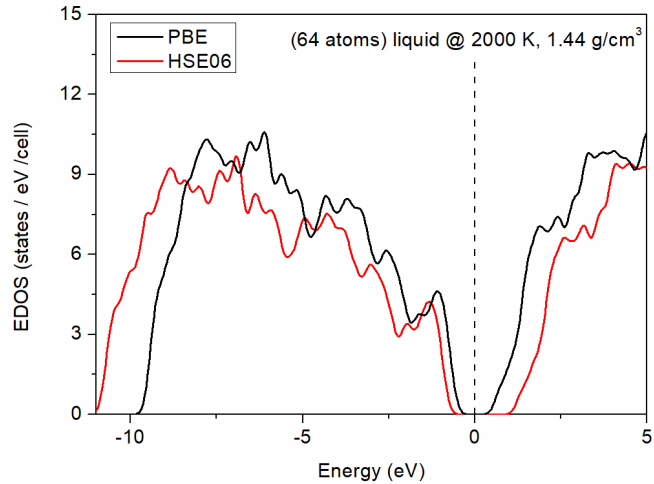

**Fig. S11.** Comparison of the electronic density of states calculated by exchange-correlation functionals of HSE06 and PBE, respectively. The Fermi level is assigned as 0 eV.

**Supplementary References**